\documentclass[a4paper]{jpconf}
\usepackage{graphicx}
\usepackage{amssymb}

\begin{document}

\title{Jointly modelling Cosmic Inflation and Dark Energy}

\author{Konstantinos Dimopoulos}

\address{Physics Department, Lancaster University, LA1 4YB, U.K.}

\ead{k.dimopoulos1@lancasaster.ac.uk}

\begin{abstract}
  Quintessential inflation utilises a single scalar field to account for the
  observations of both cosmic inflation and dark energy.
  The requirements for modelling quintessential inflation are described and two
  explicit successful models are presented in the context of
  $\alpha$-attractors and Palatini modified gravity.
\end{abstract}

\section{Introduction}

The 14~Gy history of the Universe requires special initial conditions, which
are arranged by cosmic inflation \cite{guth}. Cosmic inflation is defined
as a period of accelerated (superluminal) expansion in the early Universe.
Inflation produces a Universe that is large and uniform according to
observations.

In particle physics, inflation is typically modelled through the
inflationary paradigm, which considers that the Universe undergoes inflation
when dominated by the potential energy density of a scalar field, called the
inflaton field. The inflaton field is homogenised by the rapid expansion.

The equation of motion of a (minimally coupled, canonically normalised)
homogeneous scalar field $\phi$ is\footnote{We use natural units where
  \mbox{$c=\hbar=k_B=1$} and \mbox{$8\pi G=m_P^{-2}$}, with
  \mbox{$m_P=2.43\times 10^{18}\,$GeV} being the reduced Planck mass.}
\begin{equation}
  \ddot\phi+3H\dot\phi+V'(\phi)=0\,,
  \label{KG}
\end{equation}
where the dots denote time derivatives and the prime denotes derivative with
respect to the field. It is evident that the above equation is identical to
the one describing the roll of a body (but in field space) down its potential
$V(\phi)$ under friction due to $H$, which is the rate of the Universe
expansion (Hubble parameter). Therefore, to assist our intuition, we can
visualise the system as a ball rolling down the potential.
%as shown in Fig.~\ref{infpot}.

Since the inflationary paradigm requires the field to be potentially dominated
during inflation, its kinetic energy density should be negligible, which means
that the field slow-rolls a flat patch of $V(\phi)$, which is called the
inflationary plateau. At some point, the potential becomes steep and curved so
the kinetic energy density builds up and inflation is terminated at a critical
value $\phi_{\rm end}$.

After the end of inflation, the inflaton field rushes down to its vacuum
expectation value (VEV) and oscillates around it. These coherent oscillations
have a particle interpretation; they correspond to massive particles
(inflatons), which decay into the standard models particles that comprise the
primordial plasma. The process is called reheating.

As mentioned already, the Universe is homogenised by inflation. However, this
uniformity is not perfect. Indeed, perturbations in the density of the material
filling the Universe are necessary in order for structures (like galaxies and
galactic clusters) to eventually form. Fortunately, inflation also provides
these essential primordial density perturbations (PDPs). In the context of
the inflationary paradigm, inflation does this as follows.

%\begin{figure}
%\begin{center}
%%
%\vspace{-5cm}
%%
%\leavevmode
%\hbox{%
%\epsfxsize=4in
%%\epsffile{bicep2%
%%.eps
%\epsffile{infpot%
%.ps
%}}
%%
%\vspace{-4cm}
%%
%\begin{caption}
%\ Sketch of the inflaton scalar potential featuring an inflationary plateau,
%\end{caption}
%%
%\vspace{-.5cm}
%\label{infpot}
%\end{center}
%\end{figure}

The superluminal expansion of space during inflation amplifies the quantum
fluctuations of the inflaton field and renders them into classical
perturbations of this field, through a process called quantum decoherence.
This means that the critical value $\phi_{\rm end}$ is reached at slightly
different times at different points in space. %, as shown in Fig.~\ref{infend}.
Consequently, inflation continues a little more in some locations that in
others. This results to primordial curvature perturbations which, through the
Einstein equations, give rise to corresponding density perturbations.

Do we have any evidence of this amazing scenario? Indeed we do, because the 
PDPs reflect themselves onto the temperature perturbations of the Cosmic
Microwave Background (CMB) radiation (a kind of afterglow of the Big Bang
explosion) through the Sachs-Wolfe effect. These
perturbations have been observed at the level of $\sim 10^{-5}$. In fact there
is impressive agreement with the detailed CMB observations.
%as shown in Fig.~\ref{CMBpeaks}.
From the observations, the main characteristics of PDPs are the following.

Firstly, the PDPs are predominantly adiabatic, which means that they are
originally at the same level ($\sim 10^{-5}$) in all components of the Universe
content (neutrinos, dark matter etc.). This strongly suggests that the
primordial
plasma is comprised by the decay products of a single degree of freedom, which
could be the inflaton field. The second characteristic of the PDPs is that they
are mainly Gaussian, which reflects the inherent randomness of the inflaton's
quantum fluctuations. Finally, the PDPs are almost scale-invariant.

This approximate scale-invariance suggests that inflation is of a special type,
called quasi-de Sitter inflation, where the energy density during inflation is
almost constant. What does this translate into, in the inflationary paradigm?
Well, a homogeneous scalar field can be modelled as a perfect fluid with
barotropic parameter
\begin{equation}
w\equiv\frac{p}{\rho}=\frac{\rho_{\rm kin}-V}{\rho_{\rm kin}+V}\,,
\label{w}
\end{equation}
where $\rho_{\rm kin}\equiv\frac12\dot\phi^2$ is the kinetic energy density of the
scalar field. During inflation we have potential domination,
\mbox{$V\gg\rho_{\rm kin}$}, which means that \mbox{$w\approx -1$}. Indeed, the
expectation value of a slow-rolling inflaton is roughly constant because
ts kinetic energy density is negligible. This means that their potential energy
density remains constant so that \mbox{$\rho\simeq V\simeq\,$constant} and
inflation is quasi-de Sitter.

A lot of emphasis is put on the PDPs because they can discriminate between
different inflationary models. Apart from the PDP amplitude, which is
$\sim 10^{-5}$, there are two other observables which focus the attention of
observationalists and theorists alike. They are the spectral index of the scalar
perturbations $n_s$ and the tensor to scalar ratio $r$. At the moment,
observations suggest~\cite{planckInf}
\begin{equation}
  n_s=0.968\pm0.006\;(\mbox{1-}\sigma)\quad{\rm and}\quad
  r<0.06\;(\mbox{2-}\sigma)\,.
\label{nsr}
\end{equation}

\begin{figure}
\begin{center}
  \includegraphics{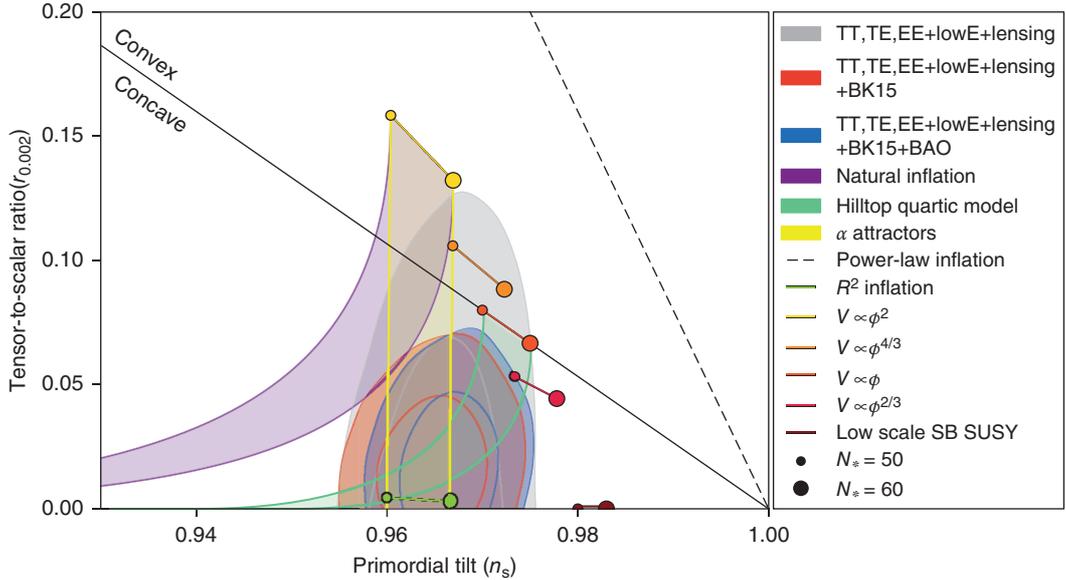}
\end{center}
\caption{\label{infobs}
 Planck sattelite observations of the spectral index $n_s$ vs the
  tensor to
  scalar ratio $r$. The blobs correspond to particular inflation models (see
  legend) considering $N=60$ [$N=50$] efolds (big [small] blob) after the
  cosmological scales exit the horizon.}
\end{figure}

In terms of the spectrum of scalar perturbations, the spectral index is %defined
\mbox{${\cal P}_\zeta(k)\propto k^{n_s(k)-1}$}, where $k$ is the momentum scale.
Notice that ${\cal P}_\zeta(k)$ is not necessarily a power-law since
$n_s=n_s(k)$. It is evident that when \mbox{$n_s=1$} then the dependence of
${\cal P}_\zeta(k)$ on $k$ disappears and the spectrum is scale-invariant.
From Eq.~(\ref{nsr}), we see that the observed values of $n_s$ are close to
unity but also significantly away from it (see also Fig,~\ref{infobs}). 

Inflation generates tensor perturbations (gravitational waves) in a similar
manner that it does scalar perturbations, which give rise to the PDP. The
ratio of the spectra of the tensor perturbations ${\cal P}_t(k)$ over the scalar
perturbations ${\cal P}_\zeta(k)$ is \mbox{$r\equiv{\cal P}_t/{\cal P}_\zeta$},
which is bounded from above at 6\%, cf.~Eq.~(\ref{nsr}). The latest
observations of $n_s$ and $r$ from the Planck satellite mission
are shown in Fig.~\ref{infobs}. The blobs in the
graph correspond to particular inflationary models. We discuss some of them
below.

\section{Examples of inflation models}

\subsection{Quartic hilltop inflation}

The name of hilltop inflation  was coined in 2005 by Lotfi Boubekeur and
David~H. Lyth \cite{hilltop}. The potential is
\begin{equation}
  V(\phi)=V_0-\lambda\phi^4+\cdots\,,
\label{hilltop}
\end{equation}
where $V_0$ is a constant density scale and \mbox{$\lambda>0$} is a
self-coupling constant, where the ellipsis denotes higher order terms, which
are needed to stabilise the potential. These terms are negligible during
inflation. The inflationary observables are analytically related as
\cite{KDhilltop}
\begin{equation}
  r=\frac83(1-n_s)\left\{1-
  \frac{\sqrt{3[2(1-n_s)N-3]}}{(1-n_s)N}\right\}\,,
\label{hilltopobs}
\end{equation}
where $N$ is the number of exponential expansions (efolds) after the
cosmological scales are pushed out of the horizon due to the superluminal
expansion during inflation. The value of $N$ depends on the reheating process.
For prompt reheating \mbox{$N\simeq 60$}, but if reheating is very inefficient
it can be decreased down to \mbox{$N\simeq 50$}. The predictions of the model
correspond to the green band in Fig.~\ref{infobs}, as specified in the legend.
It is evident that this band overlaps significantly with the 1-$\sigma$ contour
of the observations.

In order to produce the correct amplitude for the PDP, we need
\mbox{$\lambda\sim 10^{-12}$}. As a result, the VEV of the inflaton is
super-Planckian, which undermines the perturbative origin of the potential
(since all the Planck-suppressed operators become important).

\subsection{Starobinsky inflation}

This was proposed in 1980 by Alexei~A. Starobinsky \cite{staro}.
It is the first inflation model (the name was not even coined yet) and it's
fair so say that it is still the most successful one. The model is
\begin{equation}
{\cal L}=\frac12m_P^2 R+\beta R^2\,,
\label{LR2}
\end{equation}
where $R$ is the scalar curvature (Ricci scalar). The above is a modified
gravity theory. The first term on the right-hand-side is the usual
Einstein-Hilbert term. The importance of the higher order term is gauged by the
coefficient $\beta$. The theory is not a Taylor expansion, so $\beta$ is not
perturbative. Quadratic ($R^2$) gravity introduces an additional
degree of freedom which can be revealed in the form of a scalar field
(scalaron), when transforming from the modified gravity frame (Jordan frame)
to the general relativity frame (Einstein frame). This conformal transformation
redefines the metric as \mbox{$g_{\mu\nu}\rightarrow\Omega^2 g_{\mu\nu}$}, where
$\Omega^2$ is the conformal factor, given by
\mbox{$\Omega^2=\exp(\sqrt{\frac23}\,\phi/m_P)$}, where $\phi$ is the scalaron
field. In this theory, the conformal factor is
\mbox{$\Omega^2=1+4\beta R/m_P^2$}. Transforming to the Einstein frame renders
the theory as
\begin{equation}
{\cal L}=\frac12 m_P^2R+\frac12(\partial\phi)^2-V(\phi)\,,
\label{canonical}
\end{equation}  
where \mbox{$(\partial\phi)^2\equiv-\partial_\mu\phi\,\partial^\mu\phi$} and $R$
corresponds to the new metric. The potential is
\begin{equation}
  V(\phi)=\frac{m_P^4}{16\beta}\left(1-e^{-\sqrt{\frac23}\,\phi/m_P}\right)^2.
\label{Vstaro}
\end{equation}
The potential approaches a constant \mbox{$V\simeq m_P^4/16\beta$}
for large vales of the inflaton, which is generating the
inflationary plateau. The predictions of the model are
\begin{equation}
  n_s=1-\frac{2}{N}\quad{\rm and}\quad r=\frac{12}{N^2}\,.
\label{nsrstaro}
\end{equation}
As shown in Fig.~\ref{infobs}, the predictions of the model fall near the sweet
spot of the Planck observations, especially for \mbox{$N\simeq 60$}. In order to
produce the correct amplitude of PDPs we require
\mbox{$\beta=5.5225\times 10^8$}.

\subsection{$\alpha$-attractors}

A much more recent proposal generates the inflationary plateau not by
considering modified gravity, but by introducing a suitable non-minimal kinetic
term. In 2013, Renata Kallosh, Andrei Linde and Diederik Roest have introduced
the model \cite{attr}
\begin{equation}
  {\cal L}=\frac12m_P^2R+
  \frac{\frac12(\partial\varphi)^2}{\Big[1-\frac{1}{6\alpha}
\left(\frac{\varphi}{m_P}\right)^2\Big]^2}-V(\varphi)\,,
\label{Latt}
\end{equation}
where the kinetic term features poles whose location in field space is
determined by the value of the parameter \mbox{$\alpha>0$}. The above
can be obtained in supergravity with a non-trivial K\"{a}hler manifold but can
also originate in conformal theory. We can switch to a canonically normalised
field $\phi$ using the transformation
\begin{equation}
  \frac{{\rm d}\varphi}{1-\frac{1}{6\alpha}\left(\frac{\varphi}{m_P}\right)^2}
 ={\rm d}\phi\quad\Rightarrow\quad
 \frac{\varphi}{m_P}=\sqrt{6\alpha}\,\tanh\left(\frac{1}{\sqrt{6\alpha}}
 \frac{\phi}{m_P}\right)\,.
 \label{Xmation}
\end{equation}
The canonical field satisfies an equation of the form shown in
Eq.~(\ref{canonical}).

The remarkable characteristic of these models is that the inflationary
predictions are independent from the particular form of the potential as long
as it does not feature the same poles. Indeed, we have
\begin{equation}
  n_s=1-\frac{2}{N}\quad{\rm and}\quad r=\frac{12\alpha}{N^2}\,.
\label{nsralpha}
\end{equation}
The above predictions are identical with Starobinsky inflation
(cf. Eq.~(\ref{nsrstaro})) if \mbox{$\alpha=1$}. Modulating $\alpha$ we can
transpose the predictions vertically in Fig.~\ref{infobs} without affecting
the value of~$n_s$ (the yellow lines in the figure). This is the reason why
these models are called ``attractors''. Obviously, $\alpha$ cannot be too large
(\mbox{$\alpha<18$}).

\section{Quintessence as dark energy}

Observations suggest that the late Universe is also undergoing accelerated
expansion, attributed to an exotic substance called dark energy, which accounts
for almost 70\% of the Universe content at present. Dark energy
could correspond simply to a positive cosmological constant $\Lambda>0$.
However, this requires incredible fine-tuning of the order of $10^{-120}$, which
has been called ``the worst fine-tuning in physics'' (Lawrence Krauss).
Moreover, \mbox{$\Lambda>0$} violates the swampland conjectures which postulate
that there are no de Sitter vacua in the string landscape.

This is why alternative suggestions for explaining dark energy
(while assuming zero vacuum density) have been put forward.
One of the most prominent ones is quintessence \cite{quint},
so called because it is the fifth element after normal
(baryonic) matter, dark matter, photons (mainly CMB) and neutrinos. 

Quintessence is a scalar field, much like the inflaton field, slow-rolling down
a runaway potential, called quintessential tail. Thus, it can be said that the
Universe is currently engaging in a bout of late time inflation. This inflation
is also quasi-de Sitter, since the observations suggest that the barotropic
parameter for dark energy is \mbox{$w=1.006\pm0.045$} if constant or
\mbox{$w\in [-1, 0.95)$} if $w$ is variable \cite{planckParam}.

Of particular interest is thawing quintessence, which begins frozen at some
value at its quintessential tail, only to start unfreezing today when it
starts to become dominant. However, being a dynamical degree of freedom,
quintessence has to have its initial conditions explained. In particular,
for thawing quintessence, the initial frozen value must be such that the
potential density is comparable with the current energy density of the Universe.
This is called, the coincidence problem.

\section{Quintessential inflation}

One way to account for the coincidence problem of quintessence is to connect the
latter with inflation. In the end of the last century, P.~Jim~E. Peebles and
Alex Vilenkin have proposed quintessential inflation \cite{QI}, where the
quintessence field is identified with the inflaton. This is a natural idea
because both are scalar fields. It is also economic as it utilises one degree
of freedom to explain the history of the early and the late Universe. Moreover,
it models this history in a common theoretical framework. A successful model of
quintessential inflation must satisfy the observations of both cosmic inflation
and dark energy, which is rather difficult but not impossible. Finally, the
initial conditions of quintessence are fixed by the inflationary attractor, so
the coincidence problem is overcome.

Typically, in quintessential inflation the scalar potential features two flat
regions, the inflationary plateau and the quintessential tail, connected through
a sharp potential cliff. Because the inflaton field must survive until today to
become quintessence, it cannot decay into the primordial plasma, as in the
standard inflationary paradigm. Therefore, reheating has to occur by other
means. Fortunately, there are many mechanisms which can generate the primordial
radiation without the decay of the inflaton, e.g. gravitational reheating
\cite{gravreh}, instant preheating \cite{instant}, curvaton reheating
\cite{curvreh}, Ricci reheating \cite{Riccireh} and warm quintessential
inflation \cite{warmQI} to name but some.

Below two different models of Quintessential inflation are presented, to
demonstrate how such constructions are feasible in modern theory.

\subsection{Quintessential inflation with $\alpha$-attractors}

Quintessential inflation with $\alpha$-attractors has been studied in
Refs.~\cite{Charlotte,Leonora,LindeQI}.
The Lagrangian density is the one given in Eq.~(\ref{Latt}) with the potential
of the non-minimal inflaton being
\mbox{$V(\varphi)=V_0\exp(-\kappa\varphi/m_P)$}
and we also add a negative cosmological constant term given by
\mbox{$V_\Lambda=-V_0e^{\kappa\sqrt{6\alpha}}$} in order to ensure that the vacuum
density is zero (as postulated to overcome the cosmological constant problem
well before the observation of dark energy). Note that a negative cosmological 
constant is natural in string theory, which considers many anti-de Sitter vacua.

Switching to the canonical field via the transformation in
Eq.~(\ref{Xmation}) the scalar potential becomes
\begin{equation}
  V(\phi)=e^{-n}V_0
  \left\{\exp\left[n\left(1-\tanh\frac{\phi}{\sqrt{6\alpha}m_P}\right)\right]
    -1\right\}\,,
\label{Vcanon}
\end{equation}
where \mbox{$n=\kappa\sqrt{6\alpha}$}. It might not be evident, but the above
potential has the desired form.

Inflation occurs on the inflationary plateau in the limit
\mbox{$\phi\rightarrow-\infty$} where the potential is reduced to
\begin{equation}
  V(\phi)\simeq e^nV_0\left(1-2ne^{\sqrt{\frac{2}{3\alpha}}\,\phi/m_P}\right)\,,
\label{VinfOIalpha}
\end{equation}
which, when \mbox{$\alpha=1$}, is identical with Starobinsky inflation along the
inflationary plateau, as evident by comparing with Eq.~(\ref{Vstaro}).

The quintessence limit corresponds to \mbox{$\phi\rightarrow+\infty$} in which
the potential is reduced to
\begin{equation}
  V(\phi)\simeq 2n e^{-n}V_0\,e^{-\sqrt{\frac{2}{3\alpha}}\,\phi/m_P},
\label{VquintQIalpha}
\end{equation}
which is a standard exponential quintessential tail.

As shown in Ref.~\cite{Leonora}, the model works for \mbox{$\alpha\simeq 0.3$}
and \mbox{$\kappa\simeq 60$}, which means that the non-canonical inflaton in the
original exponential potential is suppressed by the scale of grand unification.

\subsection{Quintessential inflation in Palatini gravity}

\begin{figure}
\begin{center}
  %\vspace{-1cm}
\includegraphics[%height=7cm, 
  width=10cm]{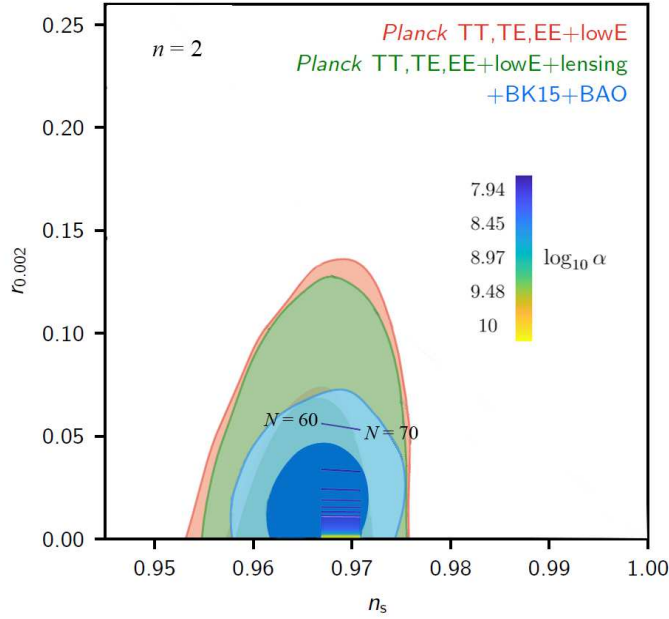}
\end{center}
\caption{\label{nsrFig}
The inflationary predictions of the model in Eq.~(\ref{VPalatini}) in
  Palatini modified gravity theory in Eq.~(\ref{LPalatini}). We see that the
  model performs very well for \mbox{$\beta\gtrsim 10^8$} (confusingly, $\beta$
  is denoted as $\alpha$ in the figure).}
\end{figure}

Recently, the framework of Palatini modified gravity has been utilised to study
quintessential inflation in Ref.~\cite{Samuel}. The Lagrangian density is
\begin{equation}
{\cal L}=\frac12m_P^2R+\beta R^2+\frac12(\partial\varphi)^2-V(\varphi)\,.
\label{LPalatini}
\end{equation}
The first part of the above is identical with the Starobinsky model in
Eq.~(\ref{LR2}), but the crucial difference is that, in Palatini gravity,
the $R^2$-term does not introduce an extra degree of freedom (no scalaron) so
that a scalar field has to be explicitly introduced, as shown in
Eq.~(\ref{LPalatini}). After a conformal transformation, we can switch to
Einstein gravity, where the Lagrangian density is
\begin{equation}
  {\cal L}=\frac12m_P^2R+
  \frac{\frac12(\partial\varphi)^2}{1+16\beta\frac{V(\varphi)}{m_P^4}}-
\frac{V(\varphi)}{1+16\beta\frac{V(\varphi)}{m_P^4}}\,.
\end{equation}
The above demonstrates that the scalar potential has become
\mbox{$U=\frac{V}{1+16\beta V/m_P^4}$}, which is very useful in designing
quintessential inflation models, because all is needed is a runaway potential
$V$ (e.g. of racetrack type). When $V$ becomes large, the unity term in the
denominator of $U$ becomes negligible and the potential $U$ plateaus with
\mbox{$U\simeq m_P^4/16\beta$}, which is exactly the Starobinsky inflationary
plateau. In the opposite limit, when $V$ is small, the denominator in $U$
approaches unity. This is also true for the kinetic term, so the field becomes
canonically normalised. Thus, any successful (thawing) quintessence model can,
in principle, work for modelling quintessential inflation in this setup,
because Palatini gravity ``flattens'' the runaway quintessence potential at
large values, creating thereby the inflationary plateau.

In Ref.~\cite{Samuel} a generalisation of the original quintessential inflation
model has been considered, where the potential is
\begin{equation}
  V(\varphi)=\left\{\begin{array}{ll}
\frac{\lambda^n}{m_P^{n-4}}(\varphi^n+M^n) & {\rm when}\;\varphi<0\\
 & \\
\frac{\lambda^n}{m_P^{n-4}}
\mbox{\large $\frac{M^{n+q}}{\varphi^q+M^q}$}
& {\rm when}\;\varphi\geq 0\,,
\end{array}\right.
\end{equation}
where $\lambda=\,$constant and $n, q$ are positive integers of order unity.
It was found that the best results are obtained when
\mbox{$(n,q)=(2,4)$}. Then, the above becomes
\begin{equation}
  V(\varphi)=\left\{\begin{array}{ll}
  m^2(\varphi^2+M^2) & {\rm when}\;\varphi<0\\
 & \\
\mbox{\large $\frac{m^2M^6}{\varphi^4+M^4}$}
& {\rm when}\;\varphi\geq 0\,,
\end{array}\right.
\label{VPalatini}
\end{equation}
where \mbox{$m=\lambda m_P$}.

When studying inflation, it was found that the correct amplitude of the PDPs
is obtained with \mbox{$m\simeq 8.8\times 10^{12}\,$GeV}. For the spectral index
and the tensor-to-scalar ratio, we find \mbox{$n_s\simeq 0.9708$} and
\mbox{$r\lesssim 0.05$}. In more detail, our findings are shown in
Fig.~\ref{nsrFig}, where it is shown that successful inflation is obtained
with \mbox{$\beta\gtrsim 10^8$}, similar to the Starobinsky values. By varying
$\beta$ one moves vertically in the $n_s-r$ graph of Fig.~\ref{nsrFig},
similarly to the case of $\alpha$-attractors (cf. Fig.~\ref{infobs}).

\begin{figure}
\begin{center}
  \vspace{-7cm}
  \centerline{\hspace{-2cm}%
    \includegraphics[%height=7cm, 
  width=15cm]{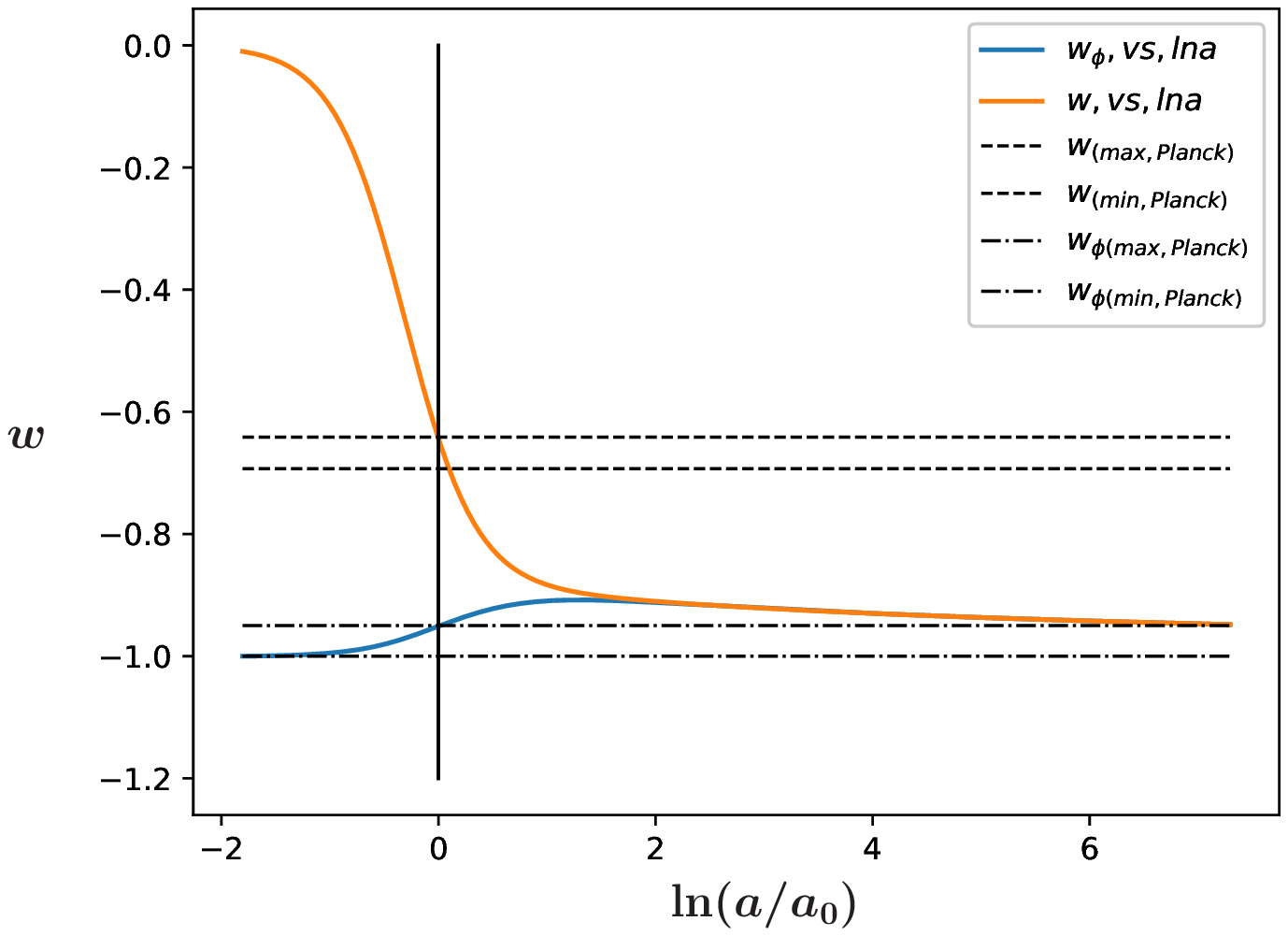}}
\vspace{-6cm}
\end{center}
\caption{\label{baroFig}
The running of the barotropic parameter of quinessence (lower curve -blue)
  and of the Universe (upper curve - orange) in the model in
  Eq.~(\ref{VPalatini}) with respect to the scale factr of the Universe.
  The vertical line corresponds to the present time, while the horizontal lines
  correspond to the observationl bounds. Originally quintessence is frozen with
  \mbox{$w_\phi=-1$}, while the Universe is matter dominated with \mbox{$w=0$}.
  As quintessence  unfreezes, its barotropic parameter $w_\phi$ grows.
  Simultaneously, quintessence begins to dominate the Universe so that the
  overal barotropic parameter $w$ decreases. In the future, quintessence fully
  dominates the Universe so \mbox{$w=w_\phi$} and the two curves merge.}
\end{figure}

When studying quintessence, we find that coincidence is attained when
\mbox{$M\sim 10\,$GeV}. No incredible fine tuning needed, in contrast to
$\Lambda$CDM. The barotropic parameter of thawing quintessence is variable.
In this limit, the model is reduced to inverse power-law quartic quintessence,
which has been studied in Ref.~\cite{Leonora}. The behaviour of the barotropic
parameters of quintessence $w_\phi$ and of the Universe $w$ is shown in
Fig.~\ref{baroFig}, plotted against the scale factor $a(t)$, which parametrises
the Universe expansion (it grows with time).

A varying dark energy barotropic parameter is parametrised as \cite{CPL}
(CPL parametrisation)
\begin{equation}
w_\phi=w_0+\left(1-\frac{a}{a_0}\right)w_a\,,
\label{CPL}
\end{equation}
which is obtained by Taylor expansion of $w_\phi(a/a_0)$ around the present time,
when \mbox{$a=a_0$}. In the above, $w_0$ and
\mbox{$w_a\equiv-({\rm d}w_\phi/{\rm d}a)_0$} are constants, soon to be measured
by forthcoming observations of the EUCLID and wFIRST (Nancy Grace Roman)
satellites. At present, the observational bounds from the Planck satellite are
\mbox{$-1\leq w_0\leq-0.95$} and \mbox{$w_a=-0.28^{+0.31}_{-0.27}$}
\cite{planckParam}.
Varying $w_0$ in the allowed region, we find
\mbox{$-0.0659<w_a<0$}, which is well in agreement with current observations and
is to be tested in the near future.

\section{Conclusions}

Cosmic inflation determines the initial conditions of the history of the
Universe and leads to a large and uniform Universe, as observed. Inflation also
generates the primordial density perturbations, which seed galaxy formation and
are reflected on the observed CMB anisotropy. The Universe today engages into a
late  inflationary period, which may be due to quintessence, a form of dark
energy. Quintessential inflation assumes that a single field drives both
inflation and quintessence, thereby allowing the study of the early and late
Universe in the context of a common theoretical framework. Quintessential
inflation leads to distinct observational signatures, such as a varying dark
energy barotropic parameter and a spike in primordial gravitational waves
\cite{sasaki} soon to be tested by observations. For more details,
see Ref.~\cite{book}.

Palatini modified gravity is a natural framework for model-building
quintessential inflation because if ``flattens'' a runaway scalar potential to
generate the desired inflationary plateau. It was demonstrated that successful
quintessential inflation in Palatini modified gravity can be achieved with a
runaway potential which interpolates between quadratic at high energies to
inverse quartic at low energies. Concrete predictions of successful Palatini
quintessential inflation may be soon tested by observations.

\section*{Acknowledgments}
This work was in (part) supported by STFC grant ST/T001038/1.


\section*{References}


\begin{thebibliography}{99}

\bibitem{guth}
Guth A H 1981 {\it Phys. Rev.} D \textbf{23} 347-56
%A.~H.~Guth,
%``The Inflationary Universe: A Possible Solution to the Horizon and Flatness Problems,''
%Phys. Rev. D \textbf{23} (1981), 347-356.
%doi:10.1103/PhysRevD.23.347

\bibitem{planckInf}
Akrami Y \textit{et al.} [Planck] 2020
{\it Astron. Astrophys.} \textbf{641} A10
%Y.~Akrami \textit{et al.} [Planck],
%``Planck 2018 results. X. Constraints on inflation,''
%Astron. Astrophys. \textbf{641} (2020), A10.
%doi:10.1051/0004-6361/201833887
%[arXiv:1807.06211 [astro-ph.CO]].

\bibitem{hilltop}
Boubekeur L and Lyth D H 2005 {\it JCAP} \textbf{07} 010
%L.~Boubekeur and D.~H.~Lyth,
%``Hilltop inflation,''
%JCAP \textbf{07} (2005), 010.
%doi:10.1088/1475-7516/2005/07/010
%[arXiv:hep-ph/0502047 [hep-ph]].

\bibitem{KDhilltop}
Dimopoulos K 2020 {\it  Phys. Lett.} B \textbf{809} 135688.
%K.~Dimopoulos,
%``An analytic treatment of quartic hilltop inflation,''
%Phys. Lett. B \textbf{809} (2020), 135688.
%doi:10.1016/j.physletb.2020.135688
%[arXiv:2006.06029 [hep-ph]].

\bibitem{staro}
Starobinsky A A 1980 {\it Phys. Lett.} B \textbf{91} 99-102
%A.~A.~Starobinsky,
%``A New Type of Isotropic Cosmological Models Without Singularity,''
%Phys. Lett. B \textbf{91} (1980), 99-102.
%doi:10.1016/0370-2693(80)90670-X

\bibitem{attr}
Kallosh R, Linde A and Roest D 2013 {\it JHEP} \textbf{11} 198
%R.~Kallosh, A.~Linde and D.~Roest,
%``Superconformal Inflationary $\alpha$-Attractors,''
%JHEP \textbf{11} (2013), 198.
%doi:10.1007/JHEP11(2013)198
%[arXiv:1311.0472 [hep-th]].

\bibitem{quint}
Caldwell R R, Dave R and Steinhardt P J 1998
{\it Phys. Rev. Lett.} \textbf{80} 1582-5
%R.~R.~Caldwell, R.~Dave and P.~J.~Steinhardt,
%``Cosmological imprint of an energy component with general equation of state,''
%Phys. Rev. Lett. \textbf{80} (1998), 1582-1585.
%doi:10.1103/PhysRevLett.80.1582
%[arXiv:astro-ph/9708069 [astro-ph]].

\bibitem{planckParam}
  Aghanim N \textit{et al.} [Planck] 2020
  {\it Astron. Astrophys.} \textbf{641} A6
%N.~Aghanim \textit{et al.} [Planck],
%``Planck 2018 results. VI. Cosmological parameters,''
%Astron. Astrophys. \textbf{641} (2020), A6.
%doi:10.1051/0004-6361/201833910
%[arXiv:1807.06209 [astro-ph.CO]].

\bibitem{QI}
Peebles J P E and Vilenkin A 1999 {\it Phys. Rev.} D \textbf{59} 063505
%P.~J.~E.~Peebles and A.~Vilenkin,
%``Quintessential inflation,''
%Phys. Rev. D \textbf{59} (1999), 063505.
%doi:10.1103/PhysRevD.59.063505
%[arXiv:astro-ph/9810509 [astro-ph]].

\bibitem{gravreh}
Ford L H 1987 {\it Phys. Rev.} D \textbf{35} 2955
%L.~H.~Ford,
%``Gravitational Particle Creation and Inflation,''
%Phys. Rev. D \textbf{35} (1987), 2955.
%doi:10.1103/PhysRevD.35.2955

\bibitem{instant}
Felder G N, Kofman L and Linde A D 1999 {\it Phys. Rev.} D \textbf{59} 123523
%G.~N.~Felder, L.~Kofman and A.~D.~Linde,
%``Instant preheating,''
%Phys. Rev. D \textbf{59} (1999), 123523.
%doi:10.1103/PhysRevD.59.123523
%[arXiv:hep-ph/9812289 [hep-ph]].

\bibitem{curvreh}
%\bibitem{Feng:2002nb}
Feng B and Li M Z 2003
{\it Phys. Lett.} B \textbf{564} 169-74
%B.~Feng and M.~z.~Li,
%``Curvaton reheating in nonoscillatory inflationary models,''
%Phys. Lett. B \textbf{564} (2003), 169-174;
%doi:10.1016/S0370-2693(03)00589-6
%[arXiv:hep-ph/0212213 [hep-ph]].
\nonum
%\bibitem{BuenoSanchez:2007jxm}
Bueno Sanchez J C and Dimopoulos K 2007 {\it JCAP} \textbf{11} 007
%J.~C.~Bueno Sanchez and K.~Dimopoulos,
%``Curvaton reheating allows TeV Hubble scale in NO inflation,''
%JCAP \textbf{11} (2007), 007.
%doi:10.1088/1475-7516/2007/11/007
%[arXiv:0707.3967 [hep-ph]].

\bibitem{Riccireh}
%\bibitem{Dimopoulos:2018wfg}
Dimopoulos K and Markkanen T 2018 {\it JCAP} \textbf{06} 021
%K.~Dimopoulos and T.~Markkanen,
%``Non-minimal gravitational reheating during kination,''
%JCAP \textbf{06} (2018), 021;
%doi:10.1088/1475-7516/2018/06/021
%[arXiv:1803.07399 [gr-qc]].
\nonum
%\bibitem{Opferkuch:2019zbd}
Opferkuch T, Schwaller P and Stefanek B A 2019 {\it JCAP} \textbf{07} 016
%T.~Opferkuch, P.~Schwaller and B.~A.~Stefanek,
%``Ricci Reheating,''
%JCAP \textbf{07} (2019), 016.
%doi:10.1088/1475-7516/2019/07/016
%[arXiv:1905.06823 [gr-qc]].

\bibitem{warmQI}
Dimopoulos K and Donaldson Wood L 2019
{\it Phys. Lett.} B \textbf{796} 26-31
%K.~Dimopoulos and L.~Donaldson-Wood,
%``Warm quintessential inflation,''
%Phys. Lett. B \textbf{796} (2019), 26-31;
%doi:10.1016/j.physletb.2019.07.017
%[arXiv:1906.09648 [gr-qc]].
\nonum
%\bibitem{Rosa:2019jci}
Rosa J G and Ventura L B 2019 {\it Phys. Lett.} B \textbf{798} 134984
%J.~G.~Rosa and L.~B.~Ventura,
%``Warm Little Inflaton becomes Dark Energy,''
%Phys. Lett. B \textbf{798} (2019), 134984.
%doi:10.1016/j.physletb.2019.134984
%[arXiv:1906.11835 [hep-ph]].

\bibitem{Charlotte}
Dimopoulos K and Owen C 2017 {\it JCAP} \textbf{06} 027
%K.~Dimopoulos and C.~Owen,
%``Quintessential Inflation with $\alpha$-attractors,''
%JCAP \textbf{06} (2017), 027.
%doi:10.1088/1475-7516/2017/06/027
%[arXiv:1703.00305 [gr-qc]].

\bibitem{Leonora}
Dimopoulos K, Donaldson Wood L and Owen C 2018
{\it Phys. Rev.} D \textbf{97} no.6 063525
%K.~Dimopoulos, L.~Donaldson Wood and C.~Owen,
%``Instant preheating in quintessential inflation with $\alpha$-attractors,''
%Phys. Rev. D \textbf{97} (2018) no.6, 063525.
%doi:10.1103/PhysRevD.97.063525
%[arXiv:1712.01760 [astro-ph.CO]].

\bibitem{LindeQI}
Akrami Y, Kallosh R, Linde A and Vardanyan V 2018 {\it JCAP} \textbf{06} 041
%Y.~Akrami, R.~Kallosh, A.~Linde and V.~Vardanyan,
%``Dark energy, $\alpha$-attractors, and large-scale structure surveys,''
%JCAP \textbf{06} (2018), 041.
%doi:10.1088/1475-7516/2018/06/041
%[arXiv:1712.09693 [hep-th]].

\bibitem{Samuel}
Dimopoulos K and S\'anchez L\'opez S 2021 
{\it Phys. Rev.} D \textbf{103} no.4 043533
%K.~Dimopoulos and S.~S\'anchez L\'opez,
%``Quintessential inflation in Palatini $f(R)$ gravity,''
%Phys. Rev. D \textbf{103} (2021) no.4, 043533.
%doi:10.1103/PhysRevD.103.043533
%[arXiv:2012.06831 [gr-qc]].
%5 citations counted in INSPIRE as of 25 Jun 2021

\bibitem{CPL}
%\bibitem{Chevallier:2000qy}
Chevallier M and Polarski D
{\it Int. J. Mod. Phys.} D \textbf{10} 213-24
%M.~Chevallier and D.~Polarski,
%``Accelerating universes with scaling dark matter,''
%Int. J. Mod. Phys. D \textbf{10} (2001), 213-224;
%doi:10.1142/S0218271801000822
%[arXiv:gr-qc/0009008 [gr-qc]].
\nonum
%\bibitem{Linder:2002et}
Linder E V 2003 {\it Phys. Rev. Lett.} \textbf{90} 091301
%E.~V.~Linder,
%``Exploring the expansion history of the universe,''
%Phys. Rev. Lett. \textbf{90} (2003), 091301.
%doi:10.1103/PhysRevLett.90.091301
%[arXiv:astro-ph/0208512 [astro-ph]].

\bibitem{sasaki}
Tashiro H, Chiba T and Sasaki M 2004
{\it Class. Quant. Grav.} \textbf{21} 1761-72
%H.~Tashiro, T.~Chiba and M.~Sasaki,
%``Reheating after quintessential inflation and gravitational waves,''
%Class. Quant. Grav. \textbf{21} (2004), 1761-1772.
%doi:10.1088/0264-9381/21/7/004
%[arXiv:gr-qc/0307068 [gr-qc]].
%96 citations counted in INSPIRE as 

\bibitem{book}
  Dimopoulos K 2021
{\it Introduction to Cosmic Inflation and Dark Energy}
(London: CRC Press/Taylor \& Francis) pp 232-43

\end{thebibliography}
\end{document}